\documentclass{aastex701}

\usepackage{subcaption} 
\shortauthors{Parashar et al. 2025}

\begin{document}
\title{Evidence for in situ particle energization during the May-2024 event based on ASPEX instrument on board Aditya-L1}

\author[orcid=0009-0004-7817-0859, gname=Shivam, sname=Parashar]{Shivam Parashar}
\affiliation{Physical Research Laboratory, Ahmedabad-380009, India}
\affiliation{Indian Institute of Technology, Gandhinagar-382355, India}
\email[show]{parasharshivam820@gmail.com}  

\author[orcid= 0000-0003-2693-5325, gname=Dibyendu, sname=Chakrabarty]{Dibyendu Chakrabarty}
\affiliation{Physical Research Laboratory, Ahmedabad-380009, India}
\email{dipu@prl.res.in}

\author[orcid=0000-0001-7590-1311,gname=Prashant, sname=Kumar]{Prashant Kumar} 
\affiliation{Physical Research Laboratory, Ahmedabad-380009, India}
\email{prashantk@prl.res.in}

\author[orcid=0009-0000-9287-6154, sname=Kumar, gname=Abhishek]{Abhishek Kumar}
\affiliation{Physical Research Laboratory, Ahmedabad-380009, India}
\email{abhishekkumar@prl.res.in}

\author[orcid=0000-0002-7248-9859, sname=Bapat, gname=Bhas]{Bhas Bapat}
\affiliation{Indian Institute of Science Education and Research, Pune-411008, India}
\email{bhas.bapat@iiserpune.ac.in}

\author[orcid=0000-0002-4781-5798, sname=Sarkar, gname=Aveek]{Aveek Sarkar}
\affiliation{Physical Research Laboratory, Ahmedabad-380009, India}
\email{aveeks@prl.res.in}

\author[orcid=0000-0003-2504-2576, sname=Janardhan, gname=P.]{P. Janardhan}
\affiliation{Physical Research Laboratory, Ahmedabad-380009, India}
\email{Janardhan.Padmanabhan@gmail.com}

\author[orcid=0000-0003-1693-453X, sname=Bhardwaj, gname=Anil]{Anil Bhardwaj}
\affiliation{Physical Research Laboratory, Ahmedabad-380009, India}
\email{bhardwaj_spl@yahoo.com}

\author[orcid=0000-0002-2050-0913, sname=Vadawale, gname=Santosh V.]{Santosh V. Vadawale}
\affiliation{Physical Research Laboratory, Ahmedabad-380009, India}
\email{santoshv@prl.res.in}

\author[sname=Shah, gname=Manan S.]{Manan S. Shah}
\affiliation{Physical Research Laboratory, Ahmedabad-380009, India}
\email{manans@prl.res.in}

\author[orcid=0000-0002-5272-6386, sname=Adalja, gname=Hiteshkumar L.]{Hiteshkumar L. Adalja}
\affiliation{Physical Research Laboratory, Ahmedabad-380009, India}
\email{hladalja@prl.res.in}

\author[sname=Patel, gname=Arpit R., orcid=0000-0002-0929-1401]{Arpit R. Patel}
\affiliation{Physical Research Laboratory, Ahmedabad-380009, India}
\email{arpitp@prl.res.in}

\author[sname=Adhyaru, gname=Pranav R.]{Pranav R. Adhyaru}
\affiliation{Physical Research Laboratory, Ahmedabad-380009, India}
\email{pranav@prl.res.in}

\author[orcid=0000-0002-5995-8681, sname=M., gname=Shanmugam]{M. Shanmugam}
\affiliation{Physical Research Laboratory, Ahmedabad-380009, India}
\email{shansm@prl.res.in}

\author[sname=Banerjee, gname=Swaroop B.]{Swaroop B. Banerjee}
\affiliation{Physical Research Laboratory, Ahmedabad-380009, India}
\email{swaroopsb@gmail.com}

\author[sname=Subramanian, gname=K. P.]{K. P. Subramanian}
\affiliation{Physical Research Laboratory, Ahmedabad-380009, India}
\email{Aniyan.Kurur@gmail.com}

\author[sname=Ladiya, gname=Tinkal, orcid=0000-0001-6022-8283]{Tinkal Ladiya}
\affiliation{Physical Research Laboratory, Ahmedabad-380009, India}
\email{tinkal@prl.res.in}

\author[sname=Sebastian, gname=Jacob, orcid=0009-0004-5904-5833]{Jacob Sebastian}
\affiliation{Physical Research Laboratory, Ahmedabad-380009, India}
\email{jacobs@prl.res.in}

\author[sname=Dalal, gname=Bijoy, orcid=0000-0001-8993-9118]{Bijoy Dalal}
\affiliation{Physical Research Laboratory, Ahmedabad-380009, India}
\email{bijoydalal.at@gmail.com}

\author[sname=Gupta, gname=Aakash, orcid=0009-0000-9917-2694]{Aakash Gupta}
\affiliation{Physical Research Laboratory, Ahmedabad-380009, India}
\affiliation{Indian Institute of Technology, Gandhinagar-382355, India}
\email{aakashgupta.du@gmail.com}

\author[orcid=0000-0002-3153-537X, sname=Goyal, gname=Shiv Kumar]{Shiv Kumar Goyal}
\affiliation{Physical Research Laboratory, Ahmedabad-380009, India}
\email{goyal@prl.res.in}

\author[orcid=0000-0003-4269-340X, sname=Tiwari, gname=Neeraj Kumar]{Neeraj Kumar Tiwari}
\affiliation{Physical Research Laboratory, Ahmedabad-380009, India}
\email{neeraj@prl.res.in}

\author[orcid=0000-0001-8900-3635, sname=Sarda, gname=Aaditya]{Aaditya Sarda}
\affiliation{Physical Research Laboratory, Ahmedabad-380009, India}
\email{aaditya@prl.res.in}

\author[sname=Kumar, gname=Sushil, orcid=0009-0004-2604-9635]{Sushil Kumar}
\affiliation{Physical Research Laboratory, Ahmedabad-380009, India}
\email{sushil@prl.res.in}

\author[orcid=0000-0002-1975-0552, sname=Singh, gname=Nishant]{Nishant Singh}
\affiliation{Physical Research Laboratory, Ahmedabad-380009, India}
\email{nishant@prl.res.in}

\author[sname=Painkra, gname=Deepak Kumar, orcid=0009-0009-8350-491X]{Deepak Kumar Painkra}
\affiliation{Physical Research Laboratory, Ahmedabad-380009, India}
\email{deepakp@prl.res.in}

\author[sname=Sharma, gname=Piyush, orcid=0000-0001-9670-1511]{Piyush Sharma}
\affiliation{Physical Research Laboratory, Ahmedabad-380009, India}
\email{piyush@prl.res.in}

\author[sname=Verma, gname=Abhishek J., orcid=0009-0009-7139-9112]{Abhishek J. Verma}
\affiliation{Physical Research Laboratory, Ahmedabad-380009, India}
\email{abhishekv@prl.res.in}

\author[sname=Dadhania, gname=M. B.]{M. B. Dadhania}
\affiliation{Physical Research Laboratory, Ahmedabad-380009, India}
\email{mbdprl@gmail.com}

\begin{abstract}

The interaction between interplanetary Coronal Mass Ejection (ICME) structures can alter the geo-effectiveness of the ICME events in myriad ways. Many aspects of these interaction processes are not well-understood till date. Using the energy spectra measured in two mutually orthogonal top hat analyzers (THA-1 and 2), which are part of the Solar Wind Ion Spectrometer (SWIS) subsystem of the Aditya Solar Wind Particle EXperiment (ASPEX) on board India’s Aditya-L1 mission, we gain insights into intricate features of ICME-ICME interactions during May 2024 solar event. We report here an unprecedented two-orthogonal-plane perspective of ICME–ICME interactions for the first time from the L1 point. The investigation reveals a special interaction region formed by the propagation of the forward shock driven by complex ejecta in the preceding ICME. The interaction causes the formation of a downstream region spanning over 13 hours, which propagates in the interplanetary medium. The observations reveal that this region serves as a site for proton and alpha particle energization, and the particles within this region get distributed from one plane to the other. The presence of forward shock and particle energization  is confirmed by the energetic particle flux measurements by the SupraThermal and Energetic Particle Spectrometer (STEPS) of ASPEX. These observations provide an unprecedented perspective on how solar wind ions become energized and distributed in an ICME-ICME interaction region.

\end{abstract}

\keywords{\uat{Solar coronal mass ejections}{310} --- \uat{Solar Wind}{1534} --- \uat{Interplanetary shocks}{829} --- \uat{Solar storm}{1526} --- \uat{Solar physics}{1476}}

\section{Introduction} \label{sec:Introduction}

During solar maximum, the rate of coronal mass ejections (CMEs) increases from a few per week to several per day, often accompanied by solar flares \citep{Hundhausen1984, Gopalswamy2004, Webb2012, Kilpua2019}. Interplanetary CMEs (ICMEs) typically propagate faster than the ambient solar wind, driving shocks observed in situ as simultaneous increases in speed and magnetic field \citep{Gopalswamy2007, Kilpua2017}. Successive CMEs may frequently erupt from the same or nearby active regions(AR) and may interact en route the Earth, producing complex structures such as multiple magnetic clouds or merged ejecta where individual ICME structures are no longer distinguishable; with prolonged magnetic field enhancement and declining speed profiles \citep{Wang2002, Burlaga2003, Lugaz2013}. ICME–ICME interactions involve reconnection, shock–shock coupling, and momentum exchange, leading to altered composition, thermal, and magnetic properties, propagation of shock waves inside the magnetic ejecta of the preceding ICME \citep{Wang2003a, Collier2007, Richardson2010, Lugaz2015b, Lugaz2016} and often enhance geo-effectiveness by intensifying southward IMF components \citep{Wang2003, Farrugia2004, Farrugia2006, Lugaz2013, Lugaz2017, Srivastava2023}. Studies by \citet{Farrugia2004} have also elucidated and reported the presence of forward and reverse shock pairs, as well as the dissipation of these shocks due to interactions of ICMEs. These studies reveal that, due to the interaction of successive ICMEs, the resulting structures may exhibit properties different from those of the participating ICMEs, such as changes in elemental abundances and charge state distributions, in addition to changes in  temperature, internal magnetic field, and pressure \citep{Lugaz2017, Srivastava2023}. Additionally, the interaction causes changes in the shape, size, orientation, and expansion rate of the ICMEs \citep{Lugaz2005, Lugaz2012, Lugaz2013, Temmer2014}.

A key aspect of ICME–ICME interactions is the variation in elemental composition and  dynamic plasma motion, particularly out of the ecliptic plane \citep{Srivastava2023}. These signatures reveal plasma mixing, shock-induced fractionation, dissipation, and kinetic-scale effects \citep{Wilson2014, Chen2018, Wilson2022, Schwartz2022, Goodrich2023}, though these effects are still not fully understood. While large-scale plasma and field parameters are routinely examined, kinetic processes such as reconnection, turbulence, and particle acceleration remain under-resolved \citep{Manchester2017}. Recently, successive eruptions from AR 13664 in May 2024 provided a unique opportunity to study such interactions. The “perfect storm” of May 2024 produced the most severe geo-effective event in two decades, with complex ejecta structures driving a prolonged storm. The Dst index reached $-412$ nT upon arrival at L1 on 10 May 2024, 16:35 UTC.
In this work, we use in situ measurements of low-energy and high-energy ion fluxes from the Solar Wind Ion Spectrometer (SWIS) and Supra-Thermal and Energetic Particle Spectrometer (STEPS) instruments of Aditya Solar Wind Particle EXperiment (ASPEX) onboard Aditya-L1 (AL1) to investigate the directional aspects of ICME–ICME interaction during the May 2024 storm. We analyze energy flux spectrograms and plasma parameters from 10–15 May 2024, enabled by the orthogonal, angularly resolved flux measurements of SWIS. Within this period, the in situ data reveal observational evidence of a special interval formed by the propagation of a shock, driven by the complex ejecta, into the preceding ICME cloud, where ion energization and energy dissipation produce an extended downstream shocked region and reveal proton and alpha particle distributions out of the ecliptic plane.
The structure of this paper is as follows. Section 2 describes the datasets and instrumentation. Section 3 presents the May 2024 event analysis, including successive ICMEs and the formation of complex ejecta structures under consideration. A detailed discussion of a special interaction interval is provided, utilizing energy flux spectrograms and directional high-energy ion fluxes from STEPS, followed by an examination of particle energization and changes in velocity distribution functions. Finally, we provide a schematic representation that summarizes the event in a nutshell. Section 4 presents the key results and conclusions.

\section{Data Set} 

For the analysis of ICME-ICME interactions for the time interval of 10 May to 15 May 2024, the observational data, i.e., solar wind bulk parameters and energy fluxes of particles, and magnetic field measurements are obtained from in situ instruments onboard Aditya-L1 \citep{Seetha2017, Tripathi2022} and Wind \citep{OGILVIE1997} spacecraft at the L1 point. The energy fluxes of proton and alpha particles in the low energy regime (0.1-20 KeV) are acquired from the SWIS instrument. Meanwhile, energetic particle fluxes for the study period are obtained from the STEPS instrument. Both these instruments are part of the ASPEX payload \citep{Janardhan2017, GOYAL2018, Goyal2025, Kumar2025} onboard the Aditya-L1 spacecraft. Henceforth, we will use AL1-ASPEX-SWIS and AL1-ASPEX-STEPS for SWIS and STEPS instruments, respectively.

Wind data include, temperature, plasma beta $(\beta)$, and velocity components $(V_x,V_y,V_z)$ from Solar Wind Experiment (SWE) \citep{Ogilvie1995} and 3DP \citep{Lin1995}, and interplanetary magnetic field (IMF) components $(B_x,B_y,B_z)$ in GSE (1-min cadence) from MFI \citep{Lepping1995}, accessed via NASA GSFC CDAWeb (\url{https://cdaweb.gsfc.nasa.gov/}). Although bulk plasma parameters are routinely derived from ASPEX observations, we deliberately use those from Wind to provide an independent confirmation of the energy histograms obtained from ASPEX in the two orthogonal planes.

AL1-ASPEX-SWIS comprises two top hat analyzers, THA-1 and THA-2, which measure the fluxes and the energy distribution of proton and alpha particles in the energy range of $0.1\text{--}20~\mathrm{KeV}$ with a time cadence of 5s.  Both the THAs have 360° angular coverage with elevation angles of $\pm 1.5^{\circ}$. THA-1 has 16 sectors with an angular resolution of $22.5^{\circ}$ each and takes observations in the sun-earth ecliptic plane. While THA-2 has 32 sectors with an angular resolution of $11.25^{\circ}$ each and takes observations in the plane perpendicular to the ecliptic plane. A detailed description of the instrument and its working is given in the paper by \citet{Kumar2025}. In this study, we use ion fluxes from sectors 9, 10, and 11 of THA-1, and sectors 15, 16, 17, 18, and 19 of THA-2. The sunward-looking sectors correspond to Sector~9 in THA-1 and Sector~17 in THA-2, with the remaining sectors oriented along the transverse directions in GSE coordinates, i.e., Sectors~9–11 (THA-1) spanning the $+Y_{\mathrm{GSE}}$ to $-Y_{\mathrm{GSE}}$ direction in the XY plane, while Sectors~15–19 (THA-2) covering the $-Z_{\mathrm{GSE}}$ to $+Z_{\mathrm{GSE}}$  direction in the XZ plane.

AL1-ASPEX-STEPS measures suprathermal and energetic particles from $20~\mathrm{keV/nucleon}$ to $6~\mathrm{MeV/nucleon}$ using six solid-state detector units; In this work, we use the data from four detectors: Intermediate (IM, FOV $15^{\circ}$), Parker Spiral (PS, FOV $30^{\circ}$), North Pointing (NP, FOV $28^{\circ}$), and Earth-Pointing (EP, FOV $37^{\circ}$) \citep{Goyal2025}. AL1-ASPEX-SWIS and AL1-ASPEX-STEPS data are available at ISSDC (\url{https://pradan.issdc.gov.in/al1/}).

\section{Event Analysis} 
 
In May 2024, the strongest geomagnetic storm of solar cycle 25 impacted the  Earth, causing a minimum Dst of $-412$ nT. It was triggered by NOAA AR 13664, one of the largest active regions (AR) in recent decades \citep{Jarolim2024}. This AR produced multiple M- and X-class flares and ICMEs \citep{Jaswal2025}. The number of ICMEs reaching the Earth varies slightly across studies, with most reporting 6–8 between 8 and 11 May \citep{Liu2024, Hayakawa2025, Thampi2025, Khuntia2025}. These works used remote-sensing observations from the Large Angle and Spectrometric Coronagraph (LASCO) onboard the Solar and Heliospheric Observatory (SOHO) \citep{Brueckner1995} and Solar-Terrestrial Relations Observatory (STEREO-A) \citep{Kaiser2008}, and in-situ data to identify CMEs linked to the ICMEs. The  ICME parameters in these studies, including flux rope tilt, angular width, and aspect ratio, were derived using the GCS forward modeling technique \citep{Thernisien2006, Thernisien2011}.
Several ICMEs erupted during 8–9 May with time gaps of $\sim$7–9 hours. They propagated towards L1, interacted, and formed Complex Ejecta 1 (CE1). Another series of ICMEs from the same AR during 9–11 May interacted to form Complex Ejecta 2 (CE2). The time gaps for CE2 were larger, $\sim$9.5–18 hours. Table 1 lists the seven ICMEs that formed CE1 and CE2. The identification and classification of these ICMEs follow earlier studies. For CE1, we infer interactions among four ICMEs (Table 1a). The other three ICMEs formed CE2 (Table 1b). ICME parameters are consistent with and taken from \citet{Liu2024}.

\begin{table}[h]
    \centering

    \begin{subtable}{\textwidth}
    \centering
    \caption*{(a) ICMEs forming CE1}
    \begin{tabular}{ccccc}
            \hline
             \rule{0pt}{3ex}
             Event & Date \& Time (UT) & Flare & 
             \shortstack{Tilt \\ Angle} & 
             \shortstack{Max \\ Speed [km/s]} \\  
            \hline
            ICME 1 & May 8, 05:09  & X1.0      & -63   &  750 \\
            ICME 2 & May 8, 12:24  & M8.7      & -55   &  850 \\
            ICME 3 & May 8, 22:36  & X1.0      & -10   & 1240 \\
            ICME 4 & May 9, 09:24  & X2.2      & -25   & 1480 \\
            \hline
        \end{tabular}
    \end{subtable}

    \vspace{0.5cm}

    \begin{subtable}{\textwidth}
        \centering
        \caption*{(b) ICMEs forming CE2}
        \begin{tabular}{ccccc}
            \hline
            \rule{0pt}{5ex}
            Event & Date \& Time (UT) & Flare & 
            \shortstack{Tilt \\ Angle } & 
            \shortstack{Max \\ Speed  [km/s]} \\
            \hline
            ICME 5 & May 9, 17:44  & X1.1 & -3  &  940 \\
            ICME 6 & May 10, 06:54 & X3.9 & 52  & 1530 \\
            ICME 7 & May 11, 01:23 & X5.8 & 58  & 1660 \\
            \hline
        \end{tabular}
    \end{subtable}

    \caption{Summary of interplanetary coronal mass ejections (ICMEs) contributing to the formation of Complex~Ejecta~1~(CE1) and Complex~Ejecta~2~(CE2).
   \textbf{(a)}~Table~1a lists the ICMEs forming CE1. 
   \textbf{(b)}~Table~1b presents the ICMEs forming CE2. 
    Both tables include parameters such as eruption time, associated flare class, flux-rope tilt angles, and maximum speeds, which are used to analyze the eruption sequence, flux-rope orientation, and interaction characteristics of the complex ejecta observed at~L1.
}

    \label{tab:icmes_ce1_ce2}
\end{table}

\begin{figure}[ht!]
\plotone{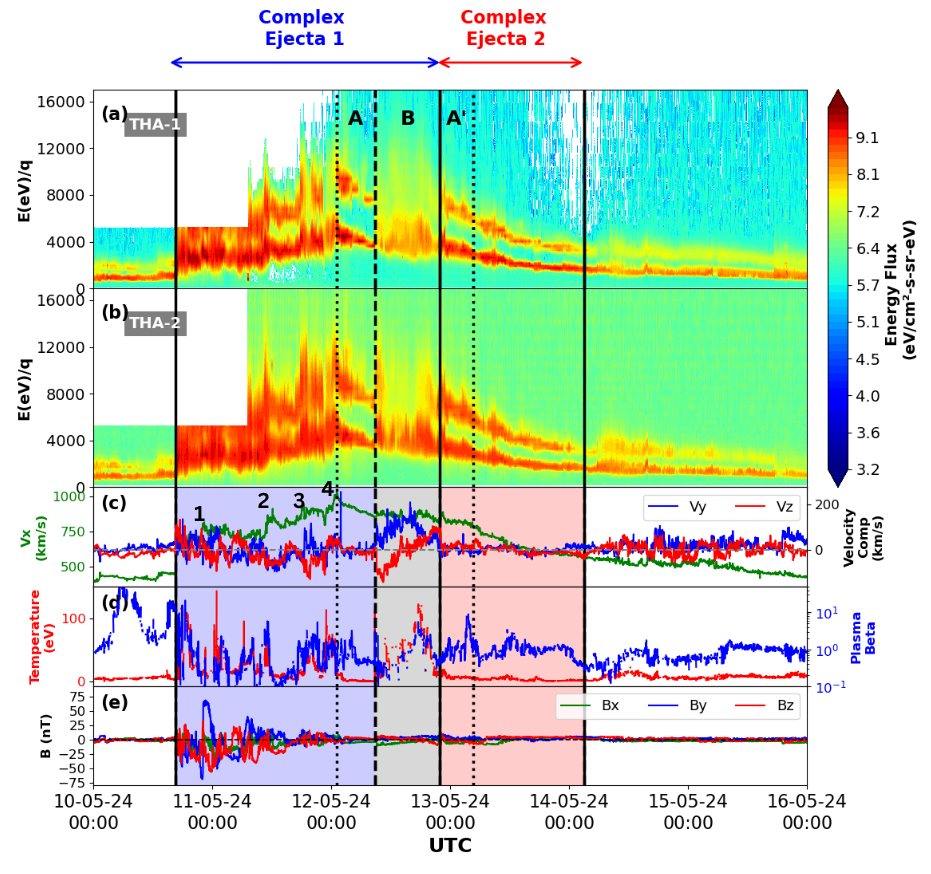}
 \caption{ Energy flux spectrograms from  AL1-ASPEX-SWIS from 10 May 2024 to 15 May 2024. The panels (a) and (b) show the energy flux spectrograms for THA-1 and THA-2 of AL1-ASPEX-SWIS. The color bar on the right of the upper two panels a and b represents the order of energy flux \(\left(\mathrm{eV}\,\mathrm{cm}^{-2}\,\mathrm{sr}^{-1}\,\mathrm{s}^{-1}\,\mathrm{eV}^{-1} \right)\), whereas the x-axis represents the time in UTC and the y-axis represents the Energy values in E(eV)/q. Panels (c) represent the velocity components $V_{x}$ (green), $V_{y}$ (blue), and $V_{z}$ (red) in km/s. Panel (d) illustrates the plasma temperature (red, in eV) and plasma beta (blue). Panels (e) represent the magnetic field components $B_{x}$ (green), $B_{y}$ (blue), and $B_{z}$ (red) in km/s. The vertical black solid lines in all the panels represent the boundaries of the complex ejecta 1 and 2, respectively. The two complex ejecta structures: Complex Ejecta 1 (CE1) comprises the UTC time interval shown by shaded blue and grey regions in panels b, c and d, while Complex Ejecta 2 (CE2) comprises the UTC time interval shown by the shaded red region in panels c, d, and e. The grey region, which is part of CE1, lies between the blue and red shaded intervals, corresponds to Interval B. The region between the first dotted line within CE1 and the dashed line represents interval~A, which is followed by interval~B~(gray~shaded). The region between the solid boundary of CE2 and the dotted line inside CE2 corresponds to interval~$A'$. The dashed line inside the CE1 represents the shock structure preceding the interval B.
}
\label{fig:fig1}
\end{figure}
In the study of \citet{Liu2024}, in situ plasma parameters were used to identify the complex ejecta formed by the interaction of ICMEs. In this study, CE1 and CE2 are marked in Figure 1. In the energy flux spectrograms shown in Figure~1, panels~(a) and~(b), the lower red traces correspond to proton fluxes (H$^+$), while the upper red traces represent alpha particle fluxes (He$^{2+}$). The energies are plotted as $E(\mathrm{eV})/q$, where $q$ denotes the particle charge, so that alpha particles appear at approximately twice the energy per charge of protons for the same velocity. The vertical black solid lines in Figure 1 mark the boundaries of two complex ejecta.
The CE1 arrives at L1 at 16:35 UTC on 10 May 2024. For the CE1, the interaction occurs as ICMEs 1-4 propagate to 1 AU. This is evident from the four bumps in the velocity profile of $V_x$ in the blue-shaded region, as seen in panel c of Figure 1. The collision between different ICME structures results in the compression of plasma, leading to an increase in temperature. From the in situ observations of the CE1, there are bumps corresponding to the interaction of ICMEs in the profile of temperature (panel d) in the blue-shaded region. During this period, there are certain intervals where the plasma beta ( the ratio of thermal pressure to  magnetic pressure) is low \citep{Richardson2010}, indicating the presence of a magnetic cloud-like structure. 
Due to the interaction, in the CE1, there is an increase in the magnetic field (B), which reaches a maximum value of $\sim$72 nT, and the peak southward $B_z$ attains the value of $\sim$59 nT in this period (see panel e in Figure 1). 
The CE2 arrives at the L1 point at 21:56 UTC on 12 May 2024 and persists until 03:00 UTC on 14 May 2024, as indicated by the red shaded region in Figure 1. This complex structure results from the interaction of three ICMEs (ICMEs 5-7 in Table 1b) events en route to the L1 point. CE2 is considerably weaker than CE1 and exhibits several features characteristic of magnetic cloud, including low plasma beta values and declining trends in velocity and temperature (see Figure 1; panels c and d). Unlike CE1, CE2 does not exhibit a significantly enhanced southward $B_z$ component. The comparatively longer gaps between the successive eruptions of ICMEs 5-7 prevented the ICMEs from merging as tightly as those forming CE1. There is a progressive shift in the magnetic flux rope tilt angles from ICMEs 1 through 7, ranging from approximately $-60^{\circ} \ \text{to} \ +60^{\circ}$ (see Table 1), further supporting the distinction between the two complex ejecta. The magnetic field strength in CE2 remains relatively low, with peak values ranging from 10 to 12 nT. The magnetic field components $(B_x, B_y, B_z)$ initially exhibit a rotation, followed by a flattening trend (not evident in the figure due to the large Y-axis scale).
Although we have followed the boundary identification of CE1 and CE2 as proposed by \citet{Liu2024}, our analysis indicates that the trailing edge of CE1 extends up to the arrival of CE2. This inference is supported by the declining velocity profiles (Figure~1 panel c) and the magnetic field configuration (Figure~1 panel e), which exhibits characteristics typical of a magnetic cloud, consistent with the trailing portion of such a structure continuing until the onset of CE2. This interpretation is further substantiated by the energy flux spectrograms discussed in the later sections, which reveal coherent variations across both orthogonal measurement planes. Moreover, the region between CE1 and CE2 ( shown in Figure 1 as a grey shaded region) — not addressed in previous studies — emerges as a distinct interval in our analysis, as evidenced through the two orthogonal-plane measurements of energy flux spectrograms (Figure~1) elaborated in the following section.

\subsection{Energy Flux Spectrograms from Two Orthogonal Planes}

The ICME–ICME interaction in the heliosphere is a complex, non-linear process that can form merged structures called complex ejecta \citep{Burlaga2002}. Past studies mainly used in situ plasma and magnetic field data to investigate these interactions \citep{Lugaz2017, Mishra2018}. Here, we analyze proton and alpha energy flux spectrograms from AL1-ASPEX-SWIS (10–15 May 2024) and directional ion fluxes from AL1-ASPEX-STEPS (10–13 May) to study ICME–ICME interactions. Energy flux, defined as the transport of $N$ particles of energy $E$ within $dE$ crossing a unit area per unit time per unit solid angle, is shown in units of $\mathrm{eV,cm^{-2}sr^{-1}s^{-1}eV^{-1}}$. The spectrograms reveal complex interaction regions and allow demarcation of substructures within the ejecta. Figure 1 shows AL1-ASPEX-SWIS energy flux spectrograms for THA-1 and THA-2. From 00:00 UTC on 10 May to 07:32 UTC on 11 May, the instrument operated up to 5 keV, later extended to 20 keV, creating a gap until 07:32 UTC on 11 May.

The arrival of CE1 is distinctly captured in the energy flux spectrogram as a shock at 16:35 UTC on 10 May, evidenced by a shift of energy fluxes to higher levels in the energy flux spectrograms of AL1-ASPEX-SWIS (Figure 1, see panels a and b) and the merging of proton and alpha populations. CE1 is characterized by a gradual increase in bulk velocity (Figure 1, panel c), with the speed profile exhibiting intermittent enhancements, suggesting ongoing interactions among the merging ICMEs as they propagate toward L1. The initial ICMEs were launched within short intervals of about $\sim$7–9 hours (Table 1(a)). This close succession likely preconditioned the ambient solar wind. As a result, the later ICMEs experienced reduced drag and were able to propagate more efficiently. Consequently, they maintained higher speeds, as indicated by the velocity increase from ~400 km/s to ~1000 km/s (Figure 1, panel c). \citet{Temmer2017} highlighted the role of transient-driven preconditioning in facilitating subsequent ICME propagation. Similarly, \citet{Kajdic2025} demonstrated through a statistical analysis that isolated ICMEs can substantially modify the interplanetary medium, enabling trailing ICMEs to decelerate less and retain higher kinetic energy over longer heliocentric distances.

Our analysis of CE1 and CE2 reveals distinct substructures within CE1 and CE2, indicating signatures of both ICME–ICME interactions and magnetic cloud–like intervals (Figure 1). We focus on this study on three specific intervals: A (12 May 01:06 UTC to 12 May 08:44 UTC), B (12 May 08:47 UTC to 12 May 21:56 UTC, and $A'$ (12 May 22:04 UTC to 13 May 04:37 UTC), as marked in Figure 1. While intervals A and $A'$ display typical magnetic cloud characteristics with low transverse flows and distinct proton–alpha separation, interval B stands out in sharp contrast. In interval B, both proton temperature and $\beta$ vary similarly (Figure 1, panel d), unlike in intervals A and $A'$. From the energy spectrogram shown in Figure 1 (panels a and b), the proton and alpha ion populations in intervals A and $A'$ appear as distinct red traces, indicating well-separated energy distributions. This suggests that the magnetic field plays a dominant role in organizing the ion populations, and these intervals are consistent with magnetic cloud–like structures. In contrast, during interval B, these populations merge and diffuse, extending to energies as high as $\sim$13-16 keV and persisting for an extended duration. Such a signature of strong interaction between ICMEs is absent during the initial arrival of CE1, despite ICME–ICME interaction already being underway. Moreover, from Figure 1 panel c, the non-radial flows ($V_y$, $V_z$) are significantly enhanced in interval B, reaching large values compared to the nearly zero flows observed in A and $A'$. These contrasting behaviors suggest that interval B demands special attention, and we investigate it in further detail in the following section.

\subsection{Interval B: Period of Enhanced Particle Energization}

Interval B, spanning from 12 May 2024, 08:47 to 12 May 2024, 21:56 UTC (Figure 1, grey-shaded interval), presents the most dynamically complex features within the interaction region, characterized by strong shock signatures, energy redistribution, and compositional anomalies in the solar wind plasma. In contrast to the WIND observations and the CFA shock database, which report the shock arrival at 09:09 UTC on 12 May, our in-situ ion fluxes from AL1-ASPEX-SWIS and AL1-ASPEX-STEPS indicate the shock arrival a little earlier, i.e., at 08:47 UTC (see Figure 1 dashed line and Figure 2a red dash-dotted line). According to the interplanetary (IP) shock database \url{https://lweb.cfa.harvard.edu/shocks/wi_data/00898/wi_00898.html}, the reduced shock speed (550.4 km/s compared to the local solar wind speed of $\sim$836 km/s) indicates that it is  driven by the leading edge of CE2. It appears to have propagated deeply into the preceding magnetic cloud associated with ICME 4. This region is characterized by a low plasma beta and compressed plasma conditions, suggesting an advanced stage of shock-cloud interaction and magnetic coalescence. Thus, this interaction leads to the formation of Interval B. The shock is quasi-perpendicular with an angle $\theta = 68.4^{\circ}$ between the upstream IMF vector and the shock normal. This shock orientation indicates a significant inclination to the radial direction from the Sun, likely due to the interaction with the preceding ICME 4. The shock has a compression ratio of 3.5, indicating it is relatively strong. As a result, there is a sudden enhancement in plasma parameters -  the magnetic field strength increases from 5 nT to 11 nT. Additionally, sudden jumps in $V_x$, $V_y$, and $V_z$ components are observed, confirming the arrival of the shock structure. As the fast and strong shock interacts with the preceding ejecta, it alters the expansion dynamics of ICME 4  \citep{Lugaz2005}. At the same time, the trailing part ICME 4 expands and flattens, attaining a nearly constant velocity of $\sim$836–900 km/s (see Figure 1, panel c). 

Most importantly, during the interval B, $\beta$ and temperature variations are in phase and the respective values are considerably higher—around 3–5 and $\sim$100 eV, respectively—indicating the dominant role of interaction as thermal effects are dominant. The magnetic field components do not exhibit significant variations during this interval.

\subsubsection{Confirmation of a Forward Shock}

\begin{figure}[ht!]
\plotone{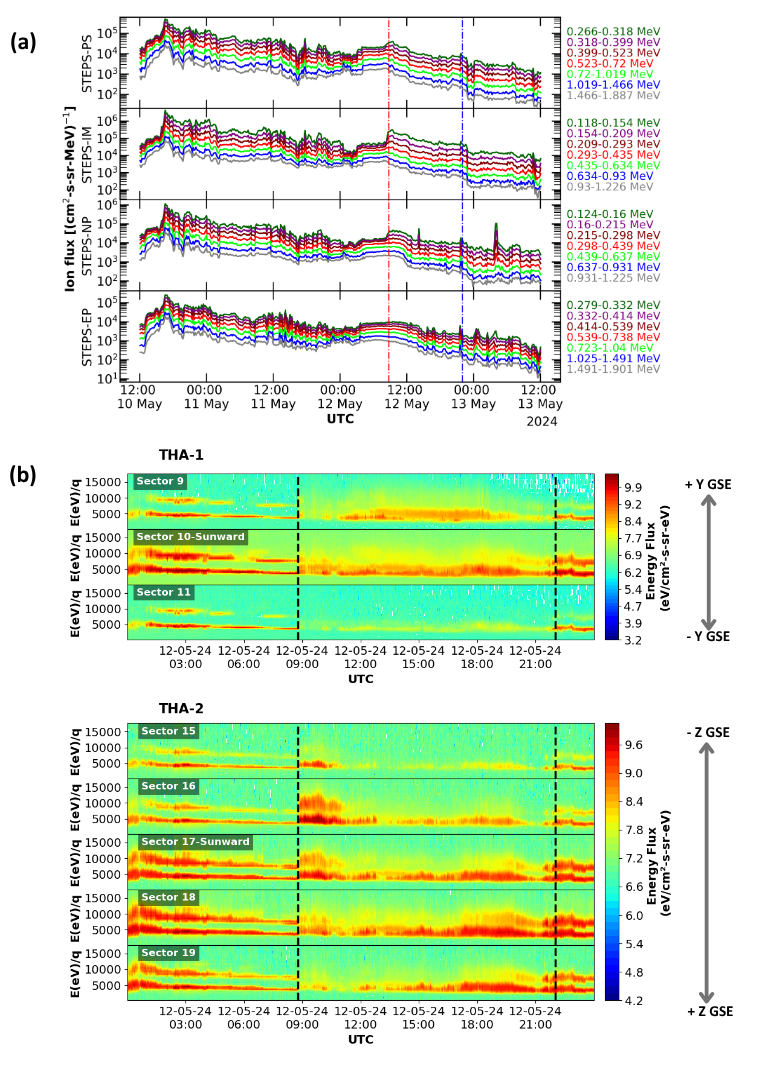}
 \caption{ (a) Energy-resolved ion flux measurements from the STEPS instrument onboard AL1-ASPEX during 10–13 May 2024. The four panels display differential ion fluxes from the STEPS-PS, STEPS-IM, STEPS-NP, and STEPS-EP detectors across energy channels ranging from $\sim$0.12 to $\sim$1.9 MeV, with color-coded lines denoting different energy bands. The interval B lies between the vertical red dash-dot (mark the forward shock arrival (08:47 UTC on 12 May) ) and blue dash-dot lines.
 (b) Time-energy spectrograms of particle fluxes measured by THA-1 (top panels) and THA-2 (bottom panels) on 12 May 2024, shown for selected angular sectors. The color scale represents the logarithm of the particle flux \([\mathrm{eV}/\mathrm{sr \cdot s \cdot cm^2 \cdot eV}]\), indicating variations in intensity as a function of energy and time. Vertical dashed lines denote boundaries encompassing Interval 3. For THA-1, Sector 10 is oriented sunward (approximately along the $-Y_{\mathrm{GSE}}$ direction), while for THA-2, Sector 17 is sunward (approximately along the $+Z_{\mathrm{GSE}}$ direction). The blue arrows indicate the approximate GSE coordinate orientation of the sectors shown in each panel, providing context for the directional information of the measured fluxes.
}
\label{fig:fig2}
\end{figure}

The presence of a forward shock and the acceleration of solar energetic particles (SEPs) is clear from Figure 2a. This figure shows energy-resolved ion flux measurements from the AL1-ASPEX-STEPS during 10–13 May 2024.  Panels in Figure 2a show time-intensity profiles at seven different energies, as indicated at the right of the figure, and also in the figure caption. The particle event shows a typical energetic storm particle (ESP) event seen previously \citep{Gosling1981, Scholer1983, Kennel1986, Desai2016}. Based on Figure 2a, a sudden enhancement in ion fluxes can be identified at 08:47 UTC on 12 May, coinciding with the arrival of a shock observed in the in situ parameters of AL1-ASPEX-SWIS. There is a gradual rise in the intensity of energetic particles prior to the crossing of the shock, followed by a gradually reducing flux downstream. Moreover, the gradual increase in the ion fluxes is seen in STEPS-PS, IM, and NP detectors. On the contrary, the STEPS-EP detector shows no shock-related signature of an increase in particle fluxes, confirming that the arrived shock is a forward shock. It is important to note that STEPS-EP is Earth-pointing and observes in the anti-sunward direction. The particle intensities peak at the shock, which is common in large  events, indicating that the shock is the primary accelerator \citep{Giacalone2012} and the acceleration occurs locally at the shock front. 
The propagation of shock leads to a nearly order-of-magnitude increase in ion flux, relative to the preceding interval before the shock. Following the shock, the downstream region exhibits a distinct change marked by gradual flux dropouts till 21:56 UTC, when a feeble but distinct signature of the arrival of a shock associated with the CE2 structure, preceded by an increase in fluxes, is seen. 

Our analysis indicates that the enhanced energetic particle fluxes observed during Interval B are primarily attributed to the shock acceleration combined with possible preconditioning effects resulting from an earlier ICME. This interpretation aligns with the findings of \citet{Gopalswamy2002}, who reported that although ICME–ICME interactions do not always produce large SEP events directly, the presence of a preceding ICME within about 12 hours of a subsequent wide and fast ICME can significantly amplify SEP production by altering the background solar wind conditions. The preconditioned medium facilitates more efficient particle acceleration at the following shock front. Moreover, \citet{Li2012} demonstrated that a preceding ICME can provide an enriched seed population and elevate turbulence levels at the trailing shock, resulting in higher acceleration efficiency and greater maximum particle energies. Together, these results substantiate our interpretation of Interval B as a region characterized by enhanced acceleration.

\subsubsection{Out-of-Plane Redistribution of Particle Fluxes}

Additional evidence pointing to the unusual plasma dynamics during this period is provided by the out-of-ecliptic redistribution of protons and alpha particles. From Figure 1, panel c, Interval B also exhibits significant enhancement in the non-radial velocity components, with noticeable fluctuations throughout. The $V_y$ component reaches a maximum of $\sim$200 km/s, while the $V_z$ component undergoes a clear rotation, transitioning from a negative maximum of $–150$ km/s to positive values up to $+100$ km/s by the end of the interval.

This indicates strong out-of-ecliptic-plane motion of the solar wind plasma, likely resulting from the propagation of a shock in the magnetic cloud of ICME 4. Simulation studies by \citet{Xiong2009}, \citet{Lugaz2012}, and \citet{Shen2012} have shown that interaction between shock driven by ICME and the preceding ICME collisions can result in significant angular deflections of up to $15^{\circ}$, especially when the ICMEs are launched with differing propagation directions. These deflections can lead to strong deviations from radial flow, consistent with the observed elevated transverse velocity components.
Supporting this interpretation, energy flux spectrograms in Figure 1 and Figure 2b reveal that THA-2—positioned to observe in the Geocentric solar ecliptic (GSE) XZ plane (perpendicular to the ecliptic)—records energy fluxes few orders of magnitude higher than those observed by THA-1, which measures in the GSE XY (ecliptic) plane. This difference is evident from the respective color scales shown in the colorbars.  Moreover, the sector-wise energy flux spectrograms of THA-1 show a clear enhancement in Sector 9 as \( V_y \) increases, suggesting a correlation between flux intensity and plasma flow direction in the \( Y_{\mathrm{GSE}} \) component. For THA-2, an initial increase in flux is observed in Sectors 15 and 16 when the \( V_z \) component is negative. Toward the end of Interval B, enhanced fluxes appear in Sectors 18 and 19, consistent with the evolution of the plasma flow direction, as indicated by the red-shaded traces in Figure 2b during Interval B. This stark contrast in flux intensities between the two planes indicates a preferential redirection of plasma out of the ecliptic. Meanwhile, it is evident that the proton fluxes are substantially higher in the THA-2 sectors compared to those of THA-1 due to interaction and distribution of the ion population. Such across-the-plane structuring has not been reported earlier for this event, highlighting the intricate dynamics introduced by the interacting transients. This behavior becomes evident only through the unique two-plane orthogonal measurements provided by the AL1–ASPEX–SWIS-THAs, offering a novel perspective on the event’s  evolution. Particularly in the case of strongly inclined, high-speed ICMEs undergoing interaction, non-radial plasma motions and directional distributions of ion populations become prominent at 1 AU. These findings underscore the importance of considering geometric effects and interaction history when interpreting solar wind structuring in the heliosphere. It may be noted that, during the interval B, the solar wind speed remains high ($\sim$ 836-900km/s, see Figure 1). However, energy flux spectrograms from THAs (Figure 1, panels a and b) indicate that the energy fluxes of protons and alphas are reduced compared to those in the preceding (blue-shaded) and the succeeding red CE2 interaction region.  Additionally, the alpha energy fluxes in THA-1 are reduced by few orders of magnitude, while THA-2 shows considerable enhancement in both proton and alpha fluxes during the interval B.
This is consistent with simulations by \citet{Lugaz2005, Lugaz2013}, who showed that interactions between ICMEs with different orientations or speeds can generate strong lateral flows and compressional effects, especially in the early stages of merging. These non-radial flows are likely driven by the differing flux rope inclinations and angular momentum redistribution resulting from the collisions \citep{Xiong2009, Shen2012}.

\subsubsection{Particle Energization: Changes in velocity distribution functions}

Interval B is formed by the propagation of a shock driven by CE2, which catches up with ICME~4 - the trailing part of CE1. The shock subsequently propagates into the rear portion of ICME~4, resulting in enhanced compression and plasma energization within this region.

Due to shock propagation, the plasma throughout this interval remains in a shocked state, with elevated proton energy fluxes at higher energies, confirming the post-shock character of the interval. 
The plasma downstream of the forward shock exhibits characteristics of shocked regions: high proton temperatures, enhanced plasma beta, and significant fluctuations in velocity components. The propagation of the forward shock into the ICME 4 (of CE1) trailing end leads to the formation of complex plasma structures in the IP medium, most notably the downstream regions (DRs)—zones of enhanced plasma compression, heating, and entropy \citep{Pitna2024, Whang1990}. In this interval, the DR existed for a duration of approximately 13 hours. During its propagation, the forward shock contributes to substantial energy dissipation, likely arising from a combination of shock heating and turbulent energy cascade that redistributes energy from across scales. Also, Turbulent heating is a key mechanism for plasma energization downstream of shocks \citep{Pitna2024}. In ICME-driven collisionless shocks, magnetic turbulence transfers energy from large-scale flows to small-scale motions, heating ions and electrons. Prior studies \citep{Smith2001, Howes2010, Matthaeus2016, Roy2022} show that turbulence-driven dissipation enhances thermal energy and produces species-dependent heating, consistent with the elevated temperatures observed during Interval B. Here, the forward shock acts as a plasma energization site, converting kinetic energy of the shock to thermal energy of the ions population.
\begin{figure}[ht!]
\plotone{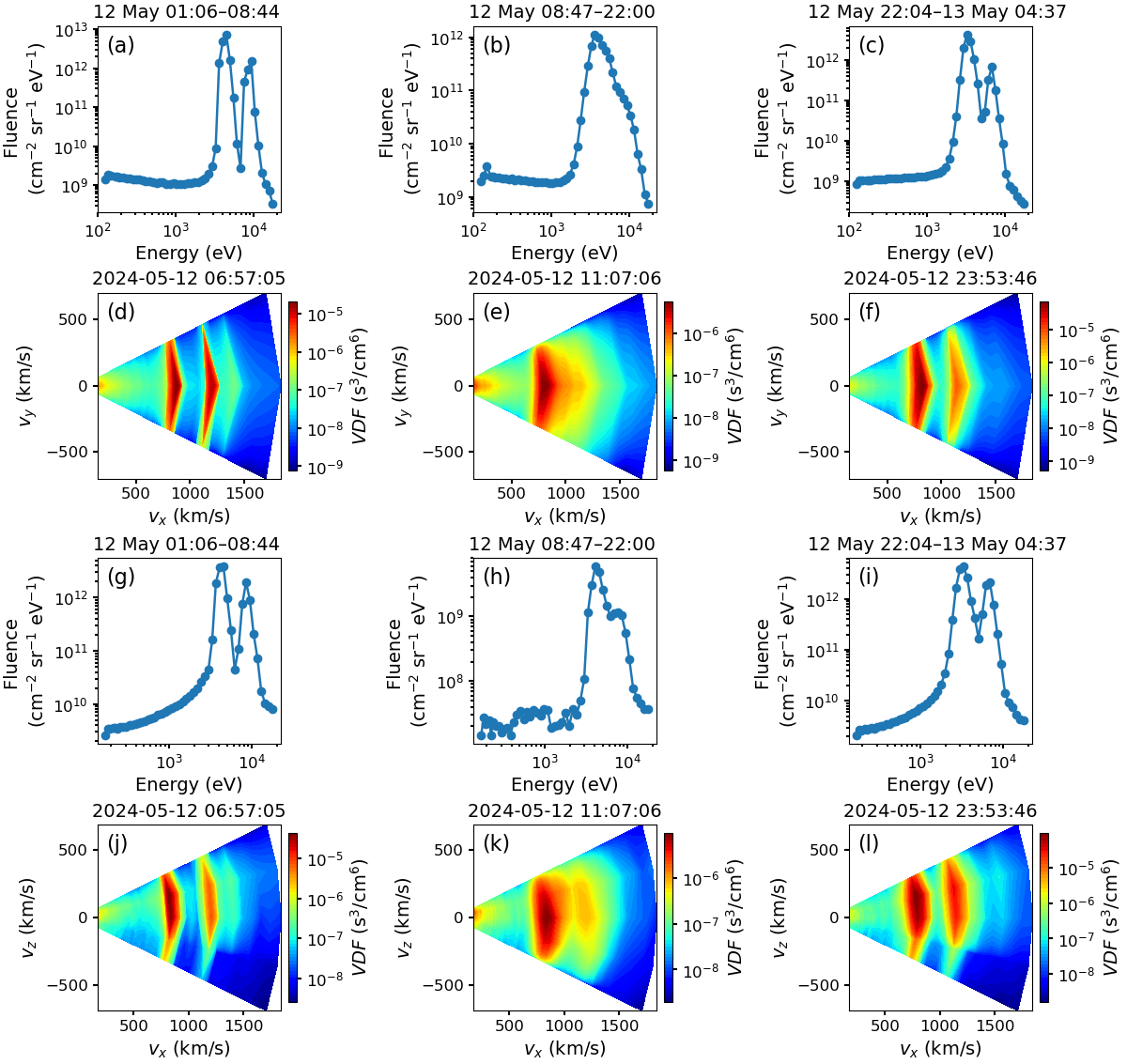}
\caption{
Temporal evolution of particle fluence spectra and velocity distribution functions (VDFs) observed by THA-1 and THA-2 during the three distinct intervals~A,~B, and~$A'$ on 12--13~May~2024. The panels in the first, middle, and rightmost columns correspond to intervals~A,~B, and~$A'$, respectively. The fluence spectra for THA-1 are shown in Figure~3, panels~(a)--(c), and for THA-2 in panels~(g)--(i), representing fluence as a function of energy~(eV) for the intervals 01:06--08:44~UTC, 08:47--22:00~UTC, and 22:04~(12~May)--04:37~(13~May)~UTC, respectively. The corresponding 2D velocity distribution functions are displayed in panels~(d)--(f) for THA-1 and panels~(j)--(l) for THA-2, shown in the $v_x$--$v_y$ and $v_x$--$v_z$ planes, respectively, with color scales indicating $VDF~[s^3/cm^6]$. The fluence plots highlight the evolution of energetic particle populations, while the VDFs reveal anisotropies and velocity-space structures that illustrate the dynamic processes of particle acceleration and transport during each interval.}

\label{fig:fig3}
\end{figure}
Kinetic processes such as ion reflection, agyrotropy, wave–particle interactions, and drift instabilities also modify energy partitioning across shocks \citep{Schwartz1998, Bale2005}. Their effectiveness depends on upstream conditions like plasma density, field strength, Mach number, magnetosonic speed, and shock geometry \citep{David2022, Schwartz2022}. In Interval B, these processes broaden particle distribution functions in velocity space, raising their second moment and producing the observed temperature increase (Figure 1, panel d).To further understand the underlying plasma dynamics, we examine ion fluence spectra and velocity distribution functions (VDFs), which provide insight into particle energization and the influence of wave–particle interactions.

Figure~3 compares the ion fluence spectra and velocity distribution functions (VDFs) for the three representative intervals~A,~B, and~$A'$ observed by THA-1 and THA-2 during 12--13~May~2024. In Figure~3, the panels in the first column correspond to Interval~A, those in the middle column to Interval~B, and the rightmost column to Interval~$A'$.
The fluence spectra are obtained by integrating the ion fluxes over the entire interval. The fluence spectra for THA-1 ( Figure 3, panels a–c) and THA-2 ( Figure 3, panels g–i) exhibit pronounced variability in both spectral slope and intensity, with significant enhancements in the keV range and extended suprathermal tails. These intervals correspond to distinct plasma regimes encountered as the solar wind traversed the complex ejecta structure. Panels (d–f) for THA-1 and (j–l) for THA-2 display VDF cuts in the (Vx, Vy) and (Vx, Vz) planes, respectively, with color-coded logarithmic scaling. VDFs are obtained using 5-minute integrated fluxes from sectors 9, 10, and 11 for THA-1, and sectors 15, 16, 17, 18, and 19 for THA-2. 

Intervals A (01:06–08:44 UT) and $A'$ (22:04–04:37 UT) are characterized by double-peaked fluence curves, indicative of distinct ion populations. Such signatures are consistent with magnetic cloud intervals, where expansion and organized magnetic field structures act to preserve multi-component populations.
Whereas Interval B (08:47–22:00 UT) exhibits a single, broad enhancement without obvious bimodality, coupled with broadened VDFs, it is indicative of plasma energization beyond thermal equilibrium. 
The corresponding VDF cuts (panels d–f, j–l) provide direct evidence of the underlying phase-space structure. Interval A (panels d and j) shows multi-lobed features with clear separation between proton and alpha populations. Interval $A'$ (panels f and l) similarly exhibits distinct components, though with broader wings compared to Interval A. By contrast, Interval B (panels e and k) is distinguished by significant broadening of the core distribution, merging of the proton and alpha populations, and loss of the double-peaked structure seen in the other two intervals. The distributions in Interval B display extended phase-space coverage in both parallel (Vx) and perpendicular (Vy, Vz) directions, consistent with enhanced plasma heating and redistribution. 
The relatively broadening and population merging in Interval B suggest that this stage of the  interaction between the forward-propagating shock and the ion fluxes of ICME 4 corresponds to the most intense heating episode. In the shocked downstream region (Interval B), strong energization broadens the core distribution and merges species into a single, heated population. Such behavior is a hallmark of collisionless shock–driven heating, where compression, reflection, and wave–particle scattering redistribute energy across velocity space \citep{ Marsch2006, Bale2009, WilsonIII2018, Verscharen2019}.
In the magnetic cloud–like intervals (A and $A'$), the proton and alpha populations retain their distinct signatures, producing bimodal fluence curves, resolved proton–alpha separation, and relatively ordered VDFs. 
These results reinforce the view that solar wind transients are governed by kinetic processes that redistribute energy across velocity space, producing anisotropies and suprathermal populations that depart markedly from classical bi-Maxwellian approximations \citep{Bale2009, Klein2016}. The May 2024 event provides direct evidence of such processes in action, with VDF signatures pointing to both localized acceleration (beam formation) and global heating (broadening and anisotropy) \citep{Bale2009, Klein2016}.

\subsubsection{The Event in a Nutshell: Schematic Representation}

\begin{figure}[ht!]
\plotone{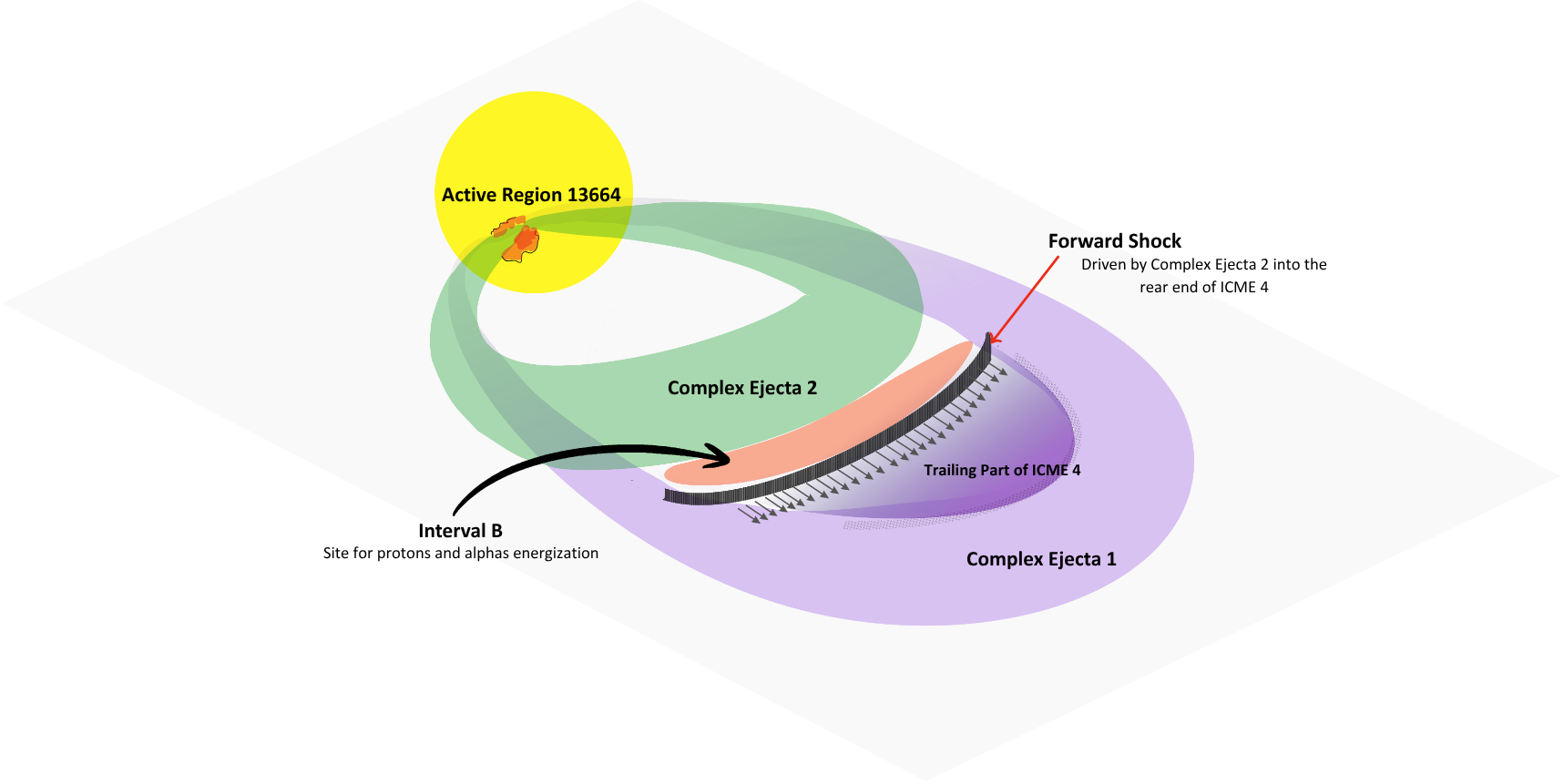}
 \caption{ 
Schematic representation of the ICME--ICME interaction sequence during May 2024. The illustration shows the propagation of Complex Ejecta~1 (CE1) and Complex Ejecta~2 (CE2) originating from Active Region~13664 on the Sun. The forward shock driven by CE2 impacts the trailing part of ICME~4 (associated with CE1), resulting in plasma compression and localized heating. 
The region labeled \textit{Interval~B} marks the site of enhanced proton and alpha-particle energization observed at L1.
}

\label{fig:fig4}
\end{figure}

Figure~\ref{fig:fig4} presents a schematic representation summarizing the sequence of events. In May 2024, successive ICMEs erupted from active region AR~13664, depicted on the Sun in Figure~\ref{fig:fig4}. These ICMEs interacted en route to form two complex ejecta structures --- CE1 (comprising ICMEs~1--4, shown in purple) and CE2 (comprising ICMEs~5--7, shown in green). CE2 is identified in the \textit{in situ} observations by its arrival at the L1 point at 21:56~UTC on 12~May.

In the schematic, CE2 drives a forward shock ahead of it, represented by a solid black line, which propagates into the trailing edge of ICME~4 (the dark purple region of CE1). The arrows indicate the direction of shock propagation. As the shock propagates into ICME~4, two key processes occur: (1) out-of-plane plasma motion, and (2) particle energization. The out-of-plane motion is evident from the enhanced $V_y$ and $V_z$ components (Figure~\ref{fig:fig1}, panel~c) in the interval following the shock and preceding the CE2 arrival. Simultaneously, as the shock propagates through ICME~4, it energizes the particles, giving rise to Interval~B represented as the red-shaded region in the schematic. This interval corresponds to a period of enhanced ion kinetic energy, where proton and alpha populations appear spread and diffused in the energy spectrograms (Figure~\ref{fig:fig1}, panels~a and~b) and in the velocity distribution functions (Figure~\ref{fig:fig3}) for the same interval.

\section{Summary and Conclusions}
\begin{enumerate}
    \item In this investigation, we have characterized different intervals (A, $A'$, and B) within the complex ejecta structures of the May 2024 ICME events with the help of in situ directional measurements of energy fluxes by  AL1-ASPEX-SWIS-THA-1 and THA-2, with accompanying bulk parameters (magnetic field, velocity components, temperature, and Plasma beta) obtained from Wind. 
    \item We have found that during Interval B, there is an across-the-plane distribution of the energy fluxes of protons and alphas in THA-2 as compared to THA-1, which is associated with high values of $V_y$ and $V_z$.
    \item We analyze a downstream shocked plasma region during Interval B, formed by the propagation and interaction of a forward shock driven by Complex ejecta 2 within the trailing magnetic cloud of ICME 4, which is part of Complex Ejecta 1. The fast-forward shock is identified in both AL1-ASPEX-SWIS and AL1-ASPEX-STEPS in situ measurements. During this interval, there is significant dispersion in the energy fluxes of protons and alpha particles, attributed to various processes associated with shock heating, turbulent cascade, and the transfer of energy to plasma species. These processes manifest as an increase in temperature and plasma beta. While previous studies have reported downstream shocked plasma regions lasting only a few seconds to minutes, in our analysis, this downstream region persisted for approximately 13 hours, which is unprecedented. Thus, these direction-resolved low and high energy observations provide direct evidence of a site for the generation of energetic particles due to the shock generated from the ICME-ICME interaction and sustaining for unusually long duration. 
\end{enumerate}

\section*{Acknowledgments}
This work is supported by the Department of Space, Government of India. The contributions from other centers of ISRO in realizing the ASPEX instrument are duly acknowledged.. We gratefully acknowledge the Wind project team for providing high-quality data through the CDAWeb platform \url{https://cdaweb.gsfc.nasa.gov/}, as well as the Principal Investigators (PIs) of the mission for their invaluable scientific contributions. We also thank the ISRO Space Science Data Center (ISSDC) for ensuring the secure storage and management of the Aditya-L1 mission data, accessible via \url{https://pradan.issdc.gov.in/al1/}.

\section*{Data Availability}
The solar wind bulk parameters and the energy fluxes for protons and alpha particles from SWIS for THA-1 and THA-2 and Ion fluxes from STEPS are accessible via \url{https://pradan1.issdc.gov.in/al1/}. The Wind spacecraft in situ parameters are obtained from CDAweb \url{https://cdaweb.gsfc.nasa.gov/}.

\bibliography{main}{}
\bibliographystyle{aasjournalv7}

\end{document}